# Dynamics of distorted skyrmions in strained chiral magnets

J. Chen[1,2], J. J. Liang[1], J. H. Yu[1], M. H. Qin[1,*], Z. Fan[1], M. Zeng[1], X. B. Lu[1], X. S. Gao[1], S. Dong[2,†], and J. -M. Liu[3]

[1]*Institute for Advanced Materials, South China Academy of Advanced Optoelectronics and Guangdong Provincial Key Laboratory of Quantum Engineering and Quantum Materials, South China Normal University, Guangzhou 510006, China*

[2]*Department of Physics, Southeast University, Nanjing 211189, China*

[3]*Laboratory of Solid State Microstructures, Nanjing University, Nanjing 210093, China*

**[Abstract]** In this work, we study the microscopic dynamics of distorted skyrmions in strained chiral magnets [K. Shibata *et al*., Nat. Nanotech. 10, 589 (2015)] under gradient magnetic field or electric current by Landau-Lifshitz-Gilbert simulations of the anisotropic spin model. It is observed that the dynamical responses are also anisotropic, and the velocities of the distorted skyrmions are periodically dependent on the directions of the external stimuli. Furthermore, in addition to the uniform motion, our work also demonstrates anti-phase harmonic vibrations of the two skyrmions in nanostripes, and the frequencies are mainly determined by the exchange anisotropy. The simulated results are well explained by Thiele theory, which may provide useful information in understanding the dynamics of the distorted skyrmions in strained chiral magnets.

* qinmh@scnu.edu.cn
† sdong@seu.edu.cn

# I. INTRODUCTION

A skyrmion is a vortex-like topological spin structure where the spins point in all directions wrapping a sphere, as schematically shown in Fig. 1(a). It has been observed experimentally in chiral magnets such as MnSi and FeGe, and attracts continuous attentions due to its interesting physics and potential applications for novel data storage devices.[1-7] In these materials, the skyrmions are stabilized by the competition of the Dzyaloshinskii-Moriya (DM) interaction and the ferromagnetic (FM) exchange interaction in the presence of an external magnetic field ($h$), as revealed in earlier works.[8-10] Importantly, it can be well controlled by an ultralow current density of a few $10^2$ A cm$^2$ which is several orders of magnitude smaller than that for magnetic domain walls. Thus, skyrmions were proposed to be promising candidates for high-density and low power consumption magnetic memories, especially considering their inherent topological stability and small lateral size. Moreover, the universal current-velocity relation of the skyrmion motion independent on impurities in chiral magnets was uncovered theoretically[11] and confirmed in the most recent experiments[12].

In addition, effective control of skyrmions has been also demonstrated using other external stimuli such as gradient magnetic/electric field and uniaxial stress.[13-16] For example, a magnetic field gradient $\nabla h$ can drive a Hall-like motion of skyrmions.[17-19] Specifically, the main velocity $v_\perp$ (perpendicular to $\nabla h$) is induced by the magnetic field gradient, and the low velocity $v_\parallel$ (parallel to $\nabla h$) is induced by the damping effect. Furthermore, significant effects of uniaxial strain on the magnetic orders in chiral magnets MnSi and FeGe have been clearly uncovered in experiments.[20-22] It is demonstrated that the uniaxial pressure can significantly modulate the temperature-region of the skyrmion lattice phase in MnSi and can tune the wave vector of the helical order at zero $h$.[23-24] Our numerical calculations of the anisotropic spin model suggest that the interaction anisotropies induced by the applied uniaxial stress play an essential role in modulating the magnetic orders.[25] More interestingly, recent experiment demonstrates that even a small anisotropic strain (~0.3%) in FeGe could induce a large deformation (~20%) of the skyrmion (as depicted in Fig. 1(b)).[26] It is suggested theoretically that the magnitude of the DM interaction is prominently modulated by the lattice distortion, resulting in the deformation of the skyrmion. Furthermore, distorted skyrmion is suggested to be stabilized in thin films of chiral magnets under tilted magnetic fields and/or with anisotropic environments.[27-28]

While the deformed skyrmions are progressively uncovered, studies proceed in dynamic control of them. The study becomes very important from the following two viewpoints. On one hand, comparing with the axisymmetric skyrmion of which the dynamic response is isotropic, deformed skyrmion could be with a strikingly different response.[27] For example, recent numerically simulations demonstrate that in the presence of the anisotropic DM interaction, the skyrmion/antiskyrmion Hall angle strongly depends on the current direction, suggesting a new degree of freedom to manipulate skyrmions.[29] One may question that if similar anisotropic response of distorted skyrmion to other external stimuli such as gradient magnetic field is also available. The question is vital because the magnetic field manipulation can work for insulators, while the electric current only works for conductors. On the other hand, this study also helps one to understand the interaction between skyrmions which is important in modulating their arrangement and dynamics.[30-32] Generally speaking, compulsive potential between skyrmions in a system makes them to be isolated.[33,34] Interestingly, it was reported in recent work that skyrmion clusters could be formed in the helical phase of $Cu_2OSeO_3$ due to the attractive skyrmion-skyrmion interactions.[35] Thus, the effects of the interaction between distorted skyrmions on their dynamics deserve to be investigated in order to further understand the microscopic dynamics in strained chiral magnets.

In this work, we numerically study the dynamics of distorted skyrmions in chiral magnets, and demonstrate anisotropic dynamic responses to gradient magnetic field or electric current. Furthermore, in addition to the uniform motion, anti-phase harmonic vibrations of the two skyrmions in chiral nanostripe are uncovered, and the oscillation frequency is mainly determined by the spin configurations of the skyrmions which can be tuned by applied strain. The simulated results are well explained by Thiele theory.

## II. MODEL AND METHODS

Following the earlier works,[36,37] we study the classical Heisenberg model on the two-dimensional square lattice for strained chiral magnets, and the Hamiltonian is given by

$$H = -J\sum_{\mathbf{i}}(\mathbf{S_i} \cdot \mathbf{S_{i+\hat{x}}} + \mathbf{S_i} \cdot \mathbf{S_{i+\hat{y}}})$$
$$-\sum_{\mathbf{i}}(K_x\mathbf{S_i} \times \mathbf{S_{i+\hat{x}}} \cdot \hat{x} + K_y\mathbf{S_i} \times \mathbf{S_{i+\hat{y}}} \cdot \hat{y}) - h\sum_{\mathbf{i}} S_{\mathbf{i}}^z \quad (1)$$

where $\mathbf{S_i}$ is the classical Heisenberg spin with unit length on site $\mathbf{i}$, $\hat{x}$, $\hat{y}$ are the basis vectors of the square lattice. The first term is the isotropic FM exchange interaction between the nearest neighbors with $J = 1$. The second term is the anisotropic DM interaction with a fixed $K_x = 0.5$, and the magnitude of the DM interaction anisotropy is defined by $\eta = K_y/K_x - 1$. The last term is the Zeeman coupling with $h$ applied along the [001] direction.

Similarly, we investigate the spin dynamics at zero temperature ($T$) by numerically solving the Landau-Lifshitz-Gilbert (LLG) equation:

$$\frac{d\mathbf{S_i}}{dt} = -\gamma \mathbf{S_i} \times \mathbf{f_i} + \alpha \mathbf{S_i} \times \frac{d\mathbf{S_i}}{dt}, \quad (2)$$

with the local effective field $\mathbf{f_i} = -(\partial H/\partial \mathbf{S_i})$. Here, $\gamma$ is the gyromagnetic ratio, $\alpha$ is the Gilbert damping coefficient. In this work, unless stated elsewhere, $\gamma = 1$ and $\alpha = 0.2$ are selected in reduced units, and the physical conclusion will not be affected by the values of these parameters. When a spin-polarized current is considered, the LLG equation updates to

$$\frac{d\mathbf{S_i}}{dt} = -\gamma \mathbf{S_i} \times \mathbf{f_i} + \alpha \mathbf{S_i} \times \frac{d\mathbf{S_i}}{dt} + (\mathbf{j} \cdot \nabla)\mathbf{S_i} - \beta \mathbf{S_i} \times (\mathbf{j} \cdot \nabla)\mathbf{S_i}. \quad (3)$$

The third term in the right side is the adiabatic spin-transfer-torque term describing the coupling between the spin-polarized current $\mathbf{j}$ ($v_s$ is the velocity of the conduction electrons) and localized spins, and the last $\beta$ term is the coupling owing to non-adiabatic effects.

We use the fourth-order Runge-Kutta method to solve the LLG equation. The initial spin configurations are obtained by the Monte Carlo simulations using the over-relaxation algorithm and temperature exchange method. Then, the configurations are sufficiently relaxed by solving the LLG equation. The simulation is performed on a 18 × 18 (16 × 40) square lattice to simulate isolated (multiple) skyrmions. Furthermore, we constrain the spin at the edges by $S^z = 1$ to reduce the finite lattice size effect for the case of isolated skyrmions, while apply the periodic/free boundary condition along the direction perpendicular to/along the field gradient for the case of multiple skyrmions. Subsequently, the spin dynamics are investigated under gradient fields, and the simulated results are further explained using the approach proposed by Thiele. Furthermore, the electric current driven motion of isolated skyrmions is also investigated with the periodic boundary condition is applied.

## III. RESULTS AND DISCUSSION

Fig. 1(a) and 1(b) give the initial spin configurations of single skyrmions under $h = 0.18$ at $\eta = 0$ and $\eta = 0.2$, respectively. The skyrmion with arbitrary rotation symmetry under isotropic condition ($\eta = 0$) is significantly distorted when the anisotropy of the DM interaction is introduced to simulate anisotropic condition caused by applied strain. Then, the motion of the skyrmion driven by gradient magnetic fields or electric currents with various directions (denoted by the angle $\theta$, as shown in Fig. 1(b)) are studied in detail.

First, we investigate the gradient-field-driven motion of single distorted skyrmion. Here, the gradient field $\nabla h$ is selected to be small enough (ranged from 0.0001 to 0.0002) to prevent destabilizing the skyrmion. The calculated $v_\parallel$ and $v_\perp$ of the skyrmions as functions of $\nabla h$ for various $\theta$ at $\eta = 0.16$ are shown in Fig. 2(a) and 2(b), respectively. Both $v_\parallel$ and $v_\perp$ increase linearly with $\nabla h$, the same as the earlier report. Furthermore, for fixed $\nabla h$ and $\theta$, $v_\perp$ is rather larger than $v_\parallel$, exhibiting a Hall-like motion behavior. Interestingly, $v_\parallel$ and $v_\perp$ are also dependent on $\theta$ for a fixed $\nabla h$, demonstrating an anisotropic dynamical response. Subsequently, the effects of the DM interaction anisotropy on the spin dynamics are investigated. Fig. 2(c) shows the simulated $v_\parallel$ as a function of $\theta$ for various $\eta$. The $v_\parallel/\nabla h$-$\theta$ curves are cosinusoidal with the period of $\pi$. Furthermore, the amplitude of the curve gradually increases with the increase of $\eta$. The simulated $v_\perp/\nabla h$-$\theta$ curves for various $\eta$ are shown in Fig. 2(d) which clearly demonstrates that $v_\perp$ is sinusoidal related with $2\theta$. Moreover, the curve significantly shifts toward the high $v_\perp$ side as $\nabla h$ increases.

As a matter of fact, the simulated results can be well explained by Thiele theory. In the continuum limit, the model Hamiltonian is updated to

$$H = \int [\frac{J}{2}(\nabla \mathbf{S})^2 - K_x(S^y \partial_x S^z - S^z \partial_x S^y)$$
$$- K_y(S^z \partial_y S^x - S^x \partial_y S^z) - hS^z]d^2r \quad , \tag{3}$$

where $S^\mu$ is the $\mu$ ($\mu = x, y, z$) component of the spin. According to the approach proposed by Thiele, one obtains

$$\mathbf{D}(\alpha \mathbf{v} - \beta \mathbf{v}_s) + N_{sk}\hat{\mathbf{z}} \times (\mathbf{v} - \beta \mathbf{v}_s) = -\gamma \frac{\partial H_{sk}}{\partial \mathbf{R}}, \tag{4}$$

where $N_{sk} = 4\pi$ is the topological charge which is not changed for distorted skyrmion, $H_{sk}$ is the energy of the system, and $\mathbf{R}$ is the position vector of the skyrmion center. The components of the dissipative force tensor $\mathbf{D}$ is given by

$$D_{ij} = \int d^2r(\partial_i \mathbf{S} \cdot \partial_j \mathbf{S}). \tag{5}$$

Here, $D_{xy} = D_{yx} = 0$ is obtained even for distorted skyrmions. Subsequently, the velocity of the distorted skyrmion under the gradient field can be calculated by

$$\begin{cases} v_\parallel = \dfrac{\alpha[(D_{yy} - D_{xx})\sin^2\theta - D_{yy}]}{\alpha\alpha D_{xx}D_{yy} + N_{sk}^2} \gamma \dfrac{\partial H_{sk}}{\partial \mathbf{R}} \\ v_\perp = \dfrac{2N_{sk} - (D_{yy} - D_{xx})\alpha\sin 2\theta}{2(D_{yy}D_{xx}\alpha\alpha + N_{sk}^2)} \gamma\hat{\mathbf{z}} \times \dfrac{\partial H_{sk}}{\partial \mathbf{R}} \end{cases}, \tag{6}$$

where $\partial H_{sk}/\partial \mathbf{R} = Q\nabla h$ with $Q = \Sigma_\mathbf{i}[1 - S_\mathbf{i}^z]$. It is easily noted that for a fixed $h$, noncollinear spin structures are favored and $\Sigma_\mathbf{i}S_\mathbf{i}^z$ is decreased when the DM interaction is enhanced. Thus, $Q$ increases as $\eta$ increases, resulting in the increase of the velocity. Fig. 3(a) shows the theoretical $v_\perp/\nabla h$-$\theta$ and $v_\parallel/\nabla h$-$\theta$ curves (dashed lines) and the corresponding simulated curves (solid dots) at $\eta = 0.16$. The theoretical result coincides well with the simulated one (the discrepancy of $v_\perp$ is less than 5%), further confirming our simulations. For example, Eq. 6 demonstrates that the velocity is independent on $\theta$ for $D_{yy} = D_{xx}$ which is available at $\eta = 0$. Furthermore, the value of $(D_{yy} - D_{xx})$ quickly increases with the increasing $\eta$, as clearly shown in Fig. 3(b). Thus, the magnitude of the dynamical response anisotropy (qualitatively denoted by the amplitude of the periodic modulation) is significantly increased, as revealed in our simulations. Moreover, for a fixed $\eta$, the amplitude of the curve is linearly dependent on $\alpha$, which has been confirmed in our simulations, although the corresponding results are not shown here.

For integrity, we also investigate the current-driven single distorted skyrmion motion. Based on the Thiele theory, the velocities are given by

$$\begin{cases} v_\parallel = \dfrac{(D_{yy} - D_{xx})(\alpha - \beta)N_{sk}\sin 2\theta - 2(D_{yy}D_{xx}\alpha\beta + N_{sk}^2)}{2(D_{yy}D_{xx}\alpha\alpha + N_{sk}^2)}v_s \\ v_\perp = \dfrac{(\alpha - \beta)N[(D_{yy} - D_{xx})\sin^2\theta + D_{xx}]}{D_{yy}D_{xx}\alpha\alpha + N_{sk}^2}(\hat{z} \times v_s) \end{cases} \quad (7)$$

It is clearly demonstrated that the spin dynamic response to the current is anisotropic (both $v_\parallel$ and $v_\perp$ are periodically modulated by $\theta$), and the magnitude of the response anisotropy is linearly related with $(\alpha - \beta)(D_{yy} - D_{xx})$. Fig. 4(a) and 4(b) show the LLG simulated $v_\parallel$ and $v_\perp$ as functions of $\theta$ for various $\eta$ at $\beta = 0.1$. It is noted that $(D_{yy} - D_{xx})$ quickly increases with the increasing $\eta$, resulting in the enhancement of the response anisotropy. Furthermore, for $\alpha = \beta$, $v_\parallel$ and $v_\perp$ are independent on $\theta$, as clearly shown in Fig. 4(c) and 4(d) which give the simulated results for various $\beta$ at $\eta = 0.16$. More importantly, the transverse motion of the skyrmion is eliminated, and the skyrmion propagates along the current direction with zero Hall effect. Thus, one may choose particular materials and/or apply external stimuli along particular directions to reduce $v_\parallel$ or $v_\perp$, and in turn to better control the motion of the skyrmions.

At last, we study the gradient-field-driven motions of multi-skyrmions in chiral nanostripes and pay particular attention to the effect of skyrmion-skyrmion interaction. For simplicity, the width of the nanostripe is reasonably selected to stabilize only two skyrmions as depicted in Fig. 5(a). Furthermore, $\alpha = 0$ is chosen for this case, and only Hall motions of the skyrmions are available. This conveniently allows us to focus on the effect of the skyrmion-skyrmion interaction. Fig. 5(b) gives the time-dependent $x$ components of the position-centers of the skyrmions at $\eta = 0.1$. It is clearly shown that the skyrmions propagate in a longitudinal-wave-like way. For a fixed $\eta$, the motion of each skyrmion could be decomposed into a uniform motion and a simple harmonic vibration. Furthermore, the two harmonic vibrations are with a same frequency ω, a same amplitude *A* and a fixed phase difference of π.

Moreover, it is demonstrated that ω is mainly dependent on the value of $\eta$ (Fig. 5(c)), and is hardly related to the Gilbert coefficient and the magnitude of the field gradient. Thus, it is strongly suggested that ω is the eigenfrequency determined by the configurations of the skyrmions which are modulated by $\eta$. In detail, ω increases exponentially with the increase of $\eta$, as clearly shown in Fig. 5(c). As a matter of fact, the fixed π phase difference between the two vibrations can also be explained by Thiele theory. Regarding the two skyrmions as a whole, the skyrmion-skyrmion interaction potential is independent of their central location. Thus, the average velocity of the two skyrmions will not be changed for a fixed $\eta$, resulting in the anti-phase vibrations. Furthermore, the wave vector of one of the helical orders significantly increases as $\eta$ increases, resulting in the increases of the compulsive potential between the skyrmions and of ω, as demonstrated in our simulations.

Up to now, experimental knowledge of the microscopic dynamics of distorted skyrmions in strained chiral magnets remains ambiguous. Interestingly, this work clearly demonstrates that distorted skyrmions exhibit anisotropic dynamical responses depending on the directions of the field gradient and/or applied electric current. In real materials, one may choose particular directions of external stimuli to reduce perpendicular/parallel drift velocity in order to better control the motion of the skyrmions. Moreover, in addition to the uniform motion, the anti-phase vibrations of the two skyrmions are observed, and the vibration frequency is mainly determined by the spin configurations of the skyrmions which can be modulated by applied strain. This phenomenon may provide useful information for future device design.

## IV. CONCLUSION

In conclusion, we have studied the dynamics of the distorted skyrmions in chiral magnets based on the LLG simulations of the anisotropic spin model. It is demonstrated that the velocities of the skyrmions are significantly dependent on the directions of the magnetic field gradient or the electric current, exhibiting the behavior of the anisotropic dynamical control. Furthermore, in addition to the uniform motion, the anti-phase harmonic vibrations of the two skyrmions in chiral nanostripe are observed, and whose frequency is mainly determined by the spin configurations which can be modulated by applied strain. The simulated results are well explained by Thiele theory and may provide new insights into the dynamics of the skyrmions in strained chiral magnets.

**Acknowledgements**:

The work is supported by the National Key Projects for Basic Research of China (Grant No. 2015CB921202), and the National Key Research Programme of China (Grant No. 2016YFA0300101), and the Natural Science Foundation of China (Grant No. 51332007), and the Science and Technology Planning Project of Guangdong Province (Grant No. 2015B090927006). X. Lu also thanks for the support from the project for Guangdong Province Universities and Colleges Pearl River Scholar Funded Scheme (2016).

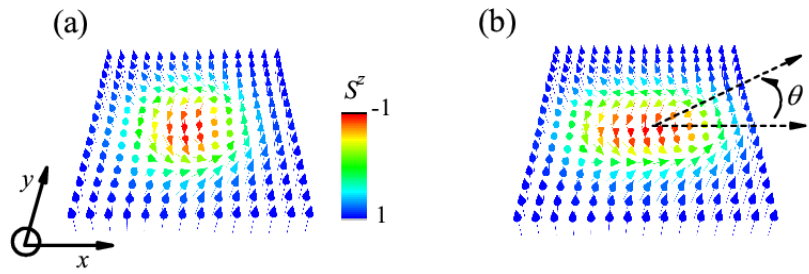

Fig.1. Spin configurations in the (a) axisymmetric skyrmion lattice phase at $\eta = 0$, and (b) distorted one at $\eta = 0.16$.

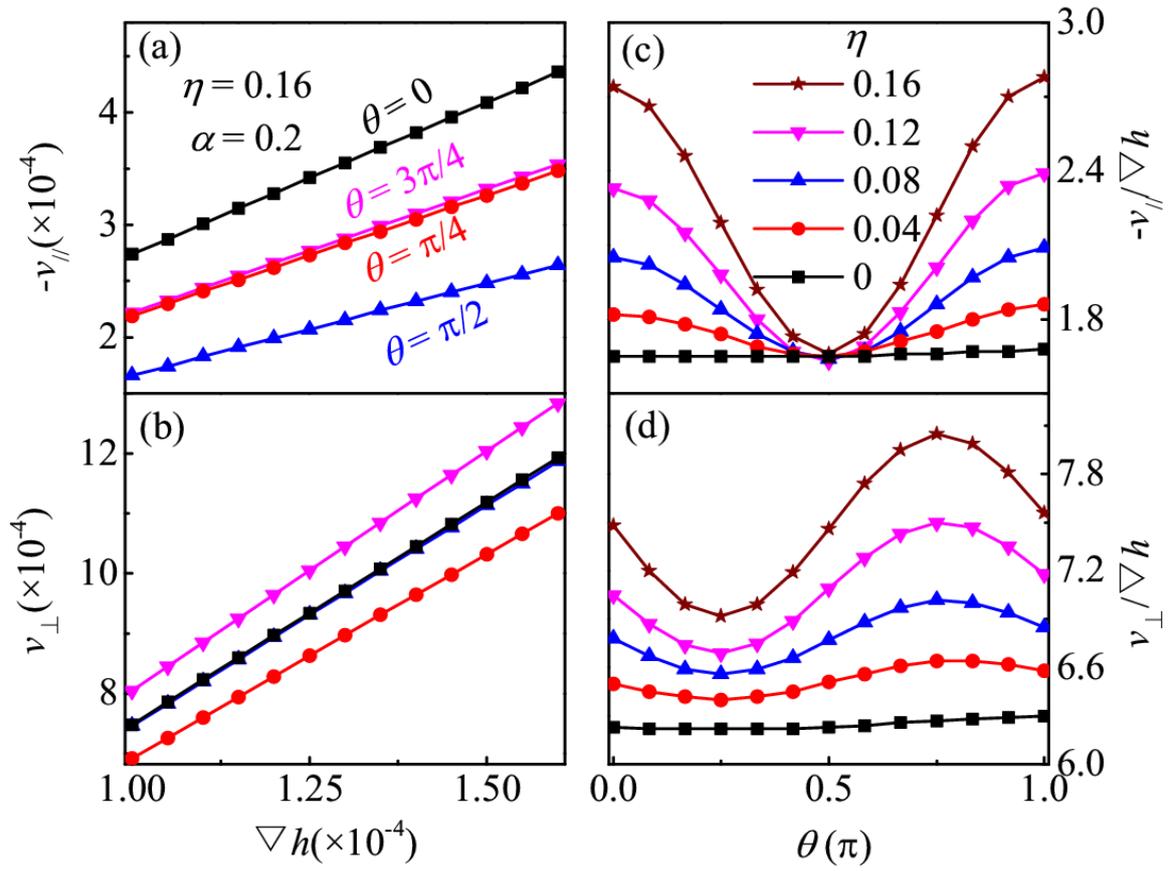

Fig.2. Magnetic field gradient driven (a) $v_{\parallel}$, and (b) $v_{\perp}$ as functions of $\nabla h$ for various $\theta$ at $\eta = 0.16$. (c) $v_{\parallel}/\nabla h$, and (d) $v_{\perp}/\nabla h$ as functions of $\theta$ for various $\eta$.

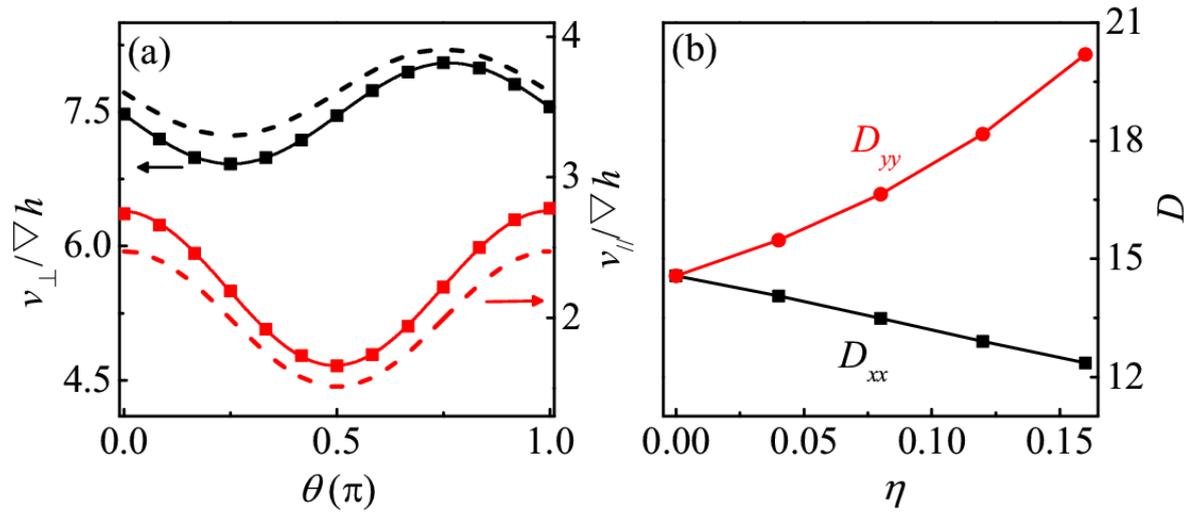

Fig.3. (a) Comparison between simulations (solid dots) and theory (dashed lines) at $\eta = 0.16$, and (b) The calculated exponents of the dissipative force tensor $D_{xx}$ and $D_{yy}$.

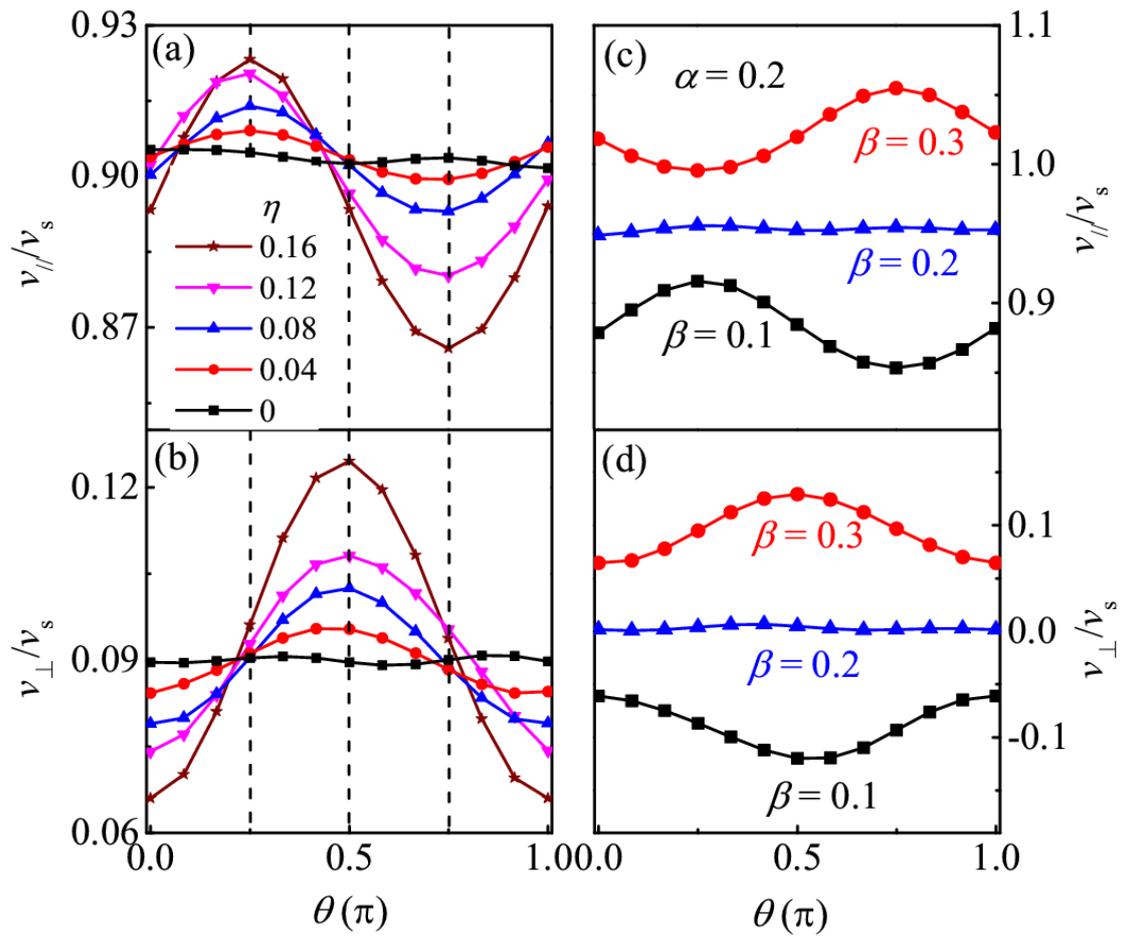

Fig.4. Electric current driven (a) $v_{\parallel}/v_s$ and (b) $v_{\perp}/v_s$ as functions of $\theta$ at $\beta = 0.1$ for various $\eta$. (c) $v_{\parallel}/v_s$ and (d) $v_{\perp}/v_s$ $v_{\perp}/v_s$ as functions of $\theta$ for various $\beta$.

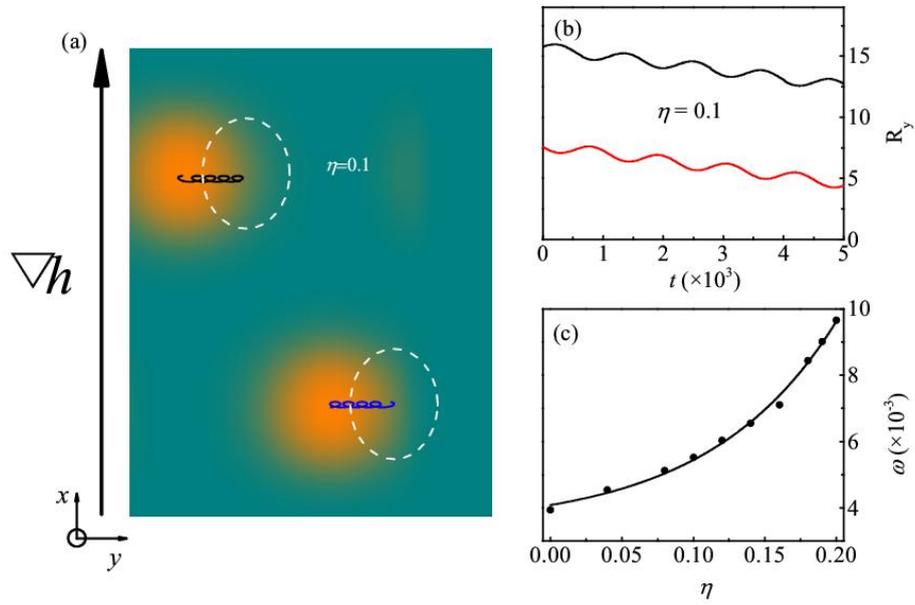

Fig.5. Magnetic field gradient driven (a) Hall motion trails for two skyrmions at $\eta = 0.1$ and $\alpha = 0$. (b) The *y*-components of the central positions of the skyrmions $R_y$ as a function of time *t*, and (c) the vibration frequency $\omega$ as a function of $\eta$.